\documentclass[showpacs,amsmath,amssymb,superscriptaddress]{revtex4}

\usepackage{graphicx}
\usepackage{dcolumn}
\usepackage{bm}

\begin{document}

\title{Manipulation of ultracold atoms in dressed adiabatic radio frequency potentials}
\author{I.~Lesanovsky}
\email{lesanovsky@atomchip.org} \affiliation{Physikalisches
Institut, Universit\"at Heidelberg, D-69120 Heidelberg, Germany}
\affiliation{Institute of Electronic Structure and Laser,
Foundation for Research and Technology - Hellas, P.O. Box
1527,GR-711 10 Heraklion, Greece}
\author{S.~Hofferberth}
\affiliation{Physikalisches Institut, Universit\"at Heidelberg,
D-69120 Heidelberg, Germany}
\author{J.~Schmiedmayer}
\affiliation{Physikalisches
Institut, Universit\"at Heidelberg, D-69120 Heidelberg, Germany}
\affiliation{Atominstitut \"Osterreichischer Universit\"aten,
TU-Wien, Vienna, Austria }
\author{Peter Schmelcher}
\email{Peter.Schmelcher@pci.uni-heidelberg.de}
\affiliation{%
Physikalisches Institut, Universit\"at Heidelberg, Philosophenweg 12, 69120 Heidelberg, Germany}%
\affiliation{%
Theoretische Chemie, Institut f\"ur Physikalische Chemie,
Universit\"at Heidelberg,
INF 229, 69120 Heidelberg, Germany}%

\date{\today}

\begin{abstract}
We explore properties of atoms whose magnetic hyperfine sub-levels
are coupled by an external magnetic radio frequency (rf) field. We
perform a thorough theoretical analysis of this driven system and
present a number of systematic approximations which eventually
give rise to dressed adiabatic radio frequency potentials. The
predictions of this analytical investigation are compared to
numerically exact results obtained by a wave packet propagation.
We outline the versatility and flexibility of this new class of
potentials and demonstrate their potential use to build atom
optical elements such as double-wells, interferometers and
ringtraps. Moreover, we perform simulations of interference
experiments carried out in rf induced double-well potentials. We
discuss how the nature of the atom-field coupling mechanism gives
rise to a decrease of the interference contrast.
\end{abstract}

\maketitle
\section{Introduction}
Using static magnetic fields in order to gain control over the
motion of neutral ground state atoms is a well-established and
frequently used experimental technique
\cite{Migdall85,Bergeman87,Schmiedmayer95,Cassettari00,Folman02}.
In particular for the manipulation of gases of atoms in the
thermal and the quantum degenerate regime magnetic fields have
been successfully employed \cite{Schreck01}. In the adiabatic
approximation the atoms are subjected to a potential which is
proportional to the modulus of the magnetic field
\cite{Folman02,Lesanovsky05}. Hence, since the field generating
structures can be almost designed at will there seems in principle
to exist a total freedom in designing potential 'landscapes'
\cite{Fortagh98,Folman02}. This virtual flexibility has resulted
in numerous proposals about how to design atom-optical elements
such as traps, guides and also interferometers
\cite{Cassettari00,Hinds01,Haensel01,Andersson02,Hommelhoff05,Esteve05}.
However, Maxwell's equations prevent magnetic traps from being
designed entirely arbitrarily. Earnshaw's theorem for instance
states that there are no local magnetic field maxima allowed to
occur inside a source-free region of space. This puts significant
constraints onto the magnetic field shapes and thus on the
trapping potentials. These restrictions can be circumvented by
superimposing time-dependent magnetic field components on the
static fields. This give rise to adiabatic potentials which show
an enormous flexibility and versatility
\cite{Zobay01,Zobay04,Schumm05,Fernholz05,Courteille05,Morizot05,Lesanovsky06}.
With the support of such time-dependent fields one can create even
complicated geometries such as rings and interferometers with
comparatively little effort. Such atom optical elements are ideal
tools to study interference \cite{Andrews97,Schumm05} and
tunneling \cite{Albiez05} of Bose-Einstein Condensates.

In this paper we discuss in detail the theoretical foundations of
the adiabatic potentials that emerge if the magnetic hyperfine
states of an alkali atom are being coupled by an external radio
frequency (rf) field. In Sec. \ref{sec:hamiltonian} we present the
underlying Hamiltonian as well as a number of transformations and
approximations which eventually give rise to the adiabatic
potentials which we present in Sec.
\ref{sec:adiabatic_potentials}. These general considerations are
followed by the discussion of a specific field configuration in
Sec. \ref{sec:ioffe_plus_rf}. We demonstrate how a very simple
setup consisting of a Ioffe-Pritchard trap \cite{Pritchard83} and
two linearly polarized rf fields allow one to create a number of
atom optical elements. In particular we outline the realization of
a tunable double well as well as a ring trap. These analytical
results are confirmed by a numerical wave packet propagation which
utilizes the exact spinor Hamiltonian. In Sec.
\ref{sec:interference} we discuss interference experiments which
are carried out by using an rf induced double-well. We outline how
the atom-field coupling which essentially forms the rf potentials
influences the outcome of interference experiments. In particular
we find an oscillation of the interference phase and a reduction
of contrast. We conclude and summarize our findings in Sec.
\ref{sec:conclusion}.

\section{General Hamiltonian of radio frequency coupled hyperfine
sub-states}\label{sec:hamiltonian} The ground state and the first
few excited states of an alkali atom are substantially split into
several hyperfine-manifolds being usually labeled by the quantum
number $F$. In the presence of a magnetic field
$\mathbf{B}(\mathbf{r},t)$ each of these manifolds again splits into
$2F+1$ branches. Let us assume the field to be of moderate strength,
i.e. not to couple adjacent hyperfine-manifolds. In this case the
dynamics of an atom within a single manifold is governed by the
Hamiltonian
\begin{eqnarray}
  H=\frac{\mathbf{p}^2}{2M}+g_F\mu_B\mathbf{F}\cdot
  \mathbf{B}(\mathbf{r},t).\label{eq:initial_hamiltonian}
\end{eqnarray}
Here $M$ is the atomic mass, $\mathbf{F}$ the $2F+1$-dimensional
representation of the angular momentum operator and $g_F$ the
corresponding $g$-factor. We assume the magnetic field to decompose
according to
\begin{eqnarray}
  \mathbf{B}(\mathbf{r},t)=\mathbf{B}_S(\mathbf{r})+\mathbf{B}_\text{RF}(\mathbf{r},t)=\mathbf{B}_S(\mathbf{r})+\sum_n
  \mathbf{B}_n(\mathbf{r})\cos(\omega t -\delta_n).
\end{eqnarray}
The static field $\mathbf{B}_S(\mathbf{r})$ serves for the purpose
of trapping, i.e. there are trapped states even in the absence of
the field $\mathbf{B}_\text{RF}(\mathbf{r},t)$ being a monochromatic
radio frequency (rf) field which is used to couple the different
magnetic hyperfine sub-states.

The static magnetic field vector $\mathbf{B}_S(\mathbf{r})$ defines
a natural quantization axis which we refer to as the $z$-axis. To
manifest this in the following we construct a unitary transformation
that rotates the angular momentum vector $\mathbf{F}$ such that it
is aligned with the local magnetic field vector. Such a
transformation which in general depends on the position $\mathbf{r}$
of the atom is given by
\begin{eqnarray}
  U_S(\mathbf{r})=\exp\left[-i{F_z}\alpha(\mathbf{r})\right]\exp\left[-i{F_y}\beta(\mathbf{r})\right]\label{eq:unitary_trafo_1}
\end{eqnarray}
with the angles
\begin{eqnarray}
\alpha(\mathbf{r})=\arctan\left[\frac{{B_S}_y(\mathbf{r})}{{B_S}_x(\mathbf{r})}\right]&\quad\text{and}\quad&
\beta(\mathbf{r})=\arctan\left[\frac{\sqrt{{B_S}_x^2(\mathbf{r})+{B_S}_y^2(\mathbf{r})}}{{B_S}_z(\mathbf{r})}\right].
\end{eqnarray}
Applying $U_S(\mathbf{r})$ to the static atom-field coupling term of
the Hamiltonian (\ref{eq:initial_hamiltonian}) yields
\begin{eqnarray}
  U_S^\dagger(\mathbf{r})\mathbf{F}U_S(\mathbf{r})\cdot
  \mathbf{B}_S(\mathbf{r})&=&\left\{\mathfrak{R}_y\left[-\beta(\mathbf{r})\right]\mathfrak{R}_z\left[-\alpha(\mathbf{r})\right]\mathbf{F}\right\}\cdot\mathbf{B}_S(\mathbf{r})\nonumber\\
 &=&\mathbf{F}\cdot\left\{\mathfrak{R}_y\left[\beta(\mathbf{r})\right]\mathfrak{R}_z\left[\alpha(\mathbf{r})\right]\mathbf{B}_S(\mathbf{r})\right\}
 =F_z\left|\mathbf{B}_S(\mathbf{r})\right|.\label{eq:rotation_of_bs}
\end{eqnarray}
As is seen here $U_S(\mathbf{r})$ indeed performs the requested
operation since the coupling to the static field has become
proportional to $F_z$. The rotation induced by $U_S(\mathbf{r})$ can
be equivalently expressed in terms of the rotation matrices
$\mathfrak{R}_i\left[\phi\right]$ acting on the vector $\mathbf{F}$.
Here $\phi$ is the rotation angle and the index $i$ denotes the axis
around which the vector is being rotated. However, instead of
rotating $\mathbf{F}$ one can equally rotate the magnetic field
vector which leads to the same result. This is exploited in the
second last equality in the series of equations
(\ref{eq:rotation_of_bs}). Consequently, applying $U_S(\mathbf{r})$
to the Hamiltonian (\ref{eq:initial_hamiltonian}) yields
\begin{eqnarray}
   U^\dagger_S(\mathbf{r})HU_S(\mathbf{r})=U^\dagger_S(\mathbf{r})\frac{\mathbf{p}^2}{2M}U_S(\mathbf{r})+g_F\mu_B F_z\left|\mathbf{B}_S(\mathbf{r})\right|+g_F\mu_B\sum_n
  \bar{\mathbf{B}}_n(\mathbf{r})\cdot\mathbf{F} \cos(\omega t
  -\delta_n)\label{eq:spatially_rotated_hamiltonian}
\end{eqnarray}
with
$\bar{\mathbf{B}}_n(\mathbf{r})=\mathfrak{R}_y\left[\beta(\mathbf{r})\right]\mathfrak{R}_z\left[\alpha(\mathbf{r})\right]\mathbf{B}_n(\mathbf{r})$.
Here the $\bar{\mathbf{B}}_n(\mathbf{r})$ are the amplitude vectors
of the rf fields now seen from a coordinate system in which the
static field $\mathbf{B}_S(\mathbf{r})$ defines the $z$-axis.

We now apply a second unitary transformation
\begin{eqnarray}
  U_R=\exp\left[-i\frac{g_F}{|g_F|}F_z\omega t\right]
  \label{eq:rotating_frame}
\end{eqnarray}
which transfers us into a frame that rotates around the $z$-axis
(local quantization axis) with the angular velocity $\omega$.
Transforming the time derivative of the Schr\"odinger equation
according to
\begin{eqnarray}
  U^\dagger_R\partial_t U_R=\partial_t+
  U^\dagger_R(\partial_tU_R)\label{eq:time_deriv_trafo}
\end{eqnarray}
one finds the effective Hamiltonian
\begin{eqnarray}
  H_\text{eff}&=&U^\dagger_RU^\dagger_S(\mathbf{r})\frac{\mathbf{p}^2}{2M}U_S(\mathbf{r})U_R+g_F\mu_B\left[\left|\mathbf{B}_S(\mathbf{r})\right|
  -\frac{g_F}{|g_F|}\frac{\hbar\omega}{g_F\mu_B}\right]F_z\nonumber\\
  &&+\frac{g_F\mu_B}{2}\sum_n\left[\left\{\left[\mathfrak{R}_z\left[-\zeta_n\right]+
  \mathfrak{R}_z\left[\zeta_n-2\frac{g_F}{|g_F|}\omega
  t\right]\right]\bar{\mathbf{B}}_n(\mathbf{r})\right\}\cdot\mathbf{F}-\bar{B_n}_z(\mathbf{r})F_z]\right]\label{eq:ham_eff_with_omega_oscillation}\\
  &&+g_F\mu_B\sum_n\bar{B_n}_z(\mathbf{r})\cos(\omega  t-\delta_n)F_z\nonumber.
\end{eqnarray}
with the phase-angle $\zeta_n=\frac{g_F}{|g_F|}\delta_n$.

In the next step we remove the last term in the Hamiltonian, which
oscillates at the frequency $\omega$. This can be done by applying
the series of unitary transformations
\begin{eqnarray}
  U_T(\mathbf{r},t)&=&\exp\left[-i\frac{g_F\mu_B}{\hbar \omega}\sum_n \bar{B_n}_z(\mathbf{r})\sin(\omega
  t-\delta_n)F_z \right]\nonumber\\
  &=&\prod_n\exp\left[-i\frac{g_F\mu_B}{\hbar \omega}\bar{B_n}_z(\mathbf{r})\sin(\omega
  t-\delta_n)F_z \right]=\prod_n
  U_{Tn}(\mathbf{r},t)\label{eq:trafo_removes_omega_term}
\end{eqnarray}
to the time-dependent Schr\"odinger equation. They have to be
carried out in accordance with equation
(\ref{eq:time_deriv_trafo}). Since $U_T(\mathbf{r},t)$ depends on
$F_z$ it does not only remove the unwanted last term of
$H_\text{eff}$ but also introduces additional ones arising from
the transformation of $F_x$ and $F_y$.  To see this we apply
$U_{Tn}(\mathbf{r},t)$ to the operator $\mathbf{F}$ which results
in
\begin{eqnarray}
  U_{Tn}^\dagger(\mathbf{r},t)\mathbf{F}U_{Tn}(\mathbf{r},t)&=&\mathfrak{R}_z\left[-\frac{g_F\mu_B}{\hbar \omega}\bar{B_z}_n(\mathbf{r})\sin(\omega
  t-\delta_n)\right]\mathbf{F}\nonumber\\
  &=&\mathfrak{R}_z\left[-\gamma_n(\mathbf{r})\sin(\omega
  t-\delta_n)\right]\mathbf{F}\\\
  &=&J_0\left[\gamma_n(\mathbf{r})\right]\mathbf{F}
  +2\sum_{m=1}^\infty J_{2m}\left[\gamma_n(\mathbf{r})\right]\cos(2 m\left[ \omega  t-\delta_n\right]) \mathbf{F}\nonumber\\
  &&+2\,\mathfrak{R}_z\left[-\frac{\pi}{2}\right]\sum_{m=0}^\infty J_{2m+1}\left[\gamma_n(\mathbf{r})\right]\sin(\left[2 m+1\right]\left[ \omega
  t-\delta_n\right])\mathbf{F}\nonumber
\end{eqnarray}
with $J_m\left[x\right]$ being the Bessel functions of the first
kind and $\gamma_n(\mathbf{r})=\frac{g_F\mu_B}{\hbar
\omega}\bar{B_n}_z(\mathbf{r})$. Since $U_T(\mathbf{r},t)$ in
general depends on the spatial coordinate the Hamiltonian which
emerges by transforming $H_\text{eff}$ reads
\begin{eqnarray}
  H_\text{RF}&=&\frac{1}{2M}\left[\mathbf{p}+\mathbf{A}(\mathbf{r},t)\right]^2+g_F\mu_B\left[\left|\mathbf{B}_S(\mathbf{r})\right|
  -\frac{g_F}{|g_F|}\frac{\hbar\omega}{g_F\mu_B}\right]F_z\label{eq:final_hamiltonian}\\
  &&+\frac{g_F\mu_B}{2}\sum_n\left[\left\{\left[\mathfrak{R}_z\left[-\zeta_n\right]+
  \mathfrak{R}_z\left[\zeta_n-2\frac{g_F}{|g_F|}\omega
  t\right]\right]\bar{\mathbf{B}}_n(\mathbf{r})\right\}\cdot
  U_{T}^\dagger(\mathbf{r},t)\mathbf{F}U_{T}(\mathbf{r},t)-\bar{B_n}_z(\mathbf{r})F_z]\right]\nonumber.
\end{eqnarray}
with the gauge potential
\begin{eqnarray}
  \mathbf{A}(\mathbf{r},t)&=& U^\dagger(\mathbf{r},t)\left(\mathbf{p} U(\mathbf{r},t)\right)\label{eq:gauge_potentials}
\end{eqnarray}
and the abbreviation
$U(\mathbf{r},t)=U_S(\mathbf{r})U_{T}(\mathbf{r},t)$ for the
combined static and time-dependent transformation.

\section{Dressed adiabatic radio frequency induced potentials}
\label{sec:adiabatic_potentials} Since we have performed only
unitary transformations the two Hamiltonians
(\ref{eq:final_hamiltonian}) and (\ref{eq:initial_hamiltonian}) are
equivalent. At first glance it is not obvious that much has been
gained since the Hamiltonian (\ref{eq:final_hamiltonian}) appears to
be extremely unhandy. However, we will show that equation
(\ref{eq:final_hamiltonian}) serves as an excellent basis for
performing a number of approximation. Finally it will allow us to
derive a time-independent expression for the dressed adiabatic
potentials.

For the first approximation we assume
\begin{eqnarray}
\gamma_n(\mathbf{r})=\left|\frac{g_F\mu_B}{\hbar
\omega}\bar{B_n}_z(\mathbf{r})\right|\ll
1,\label{eq:parallel_lamor_frequency}
\end{eqnarray}
i.e. the Lamor frequency
$\omega_{L\parallel}=\left|g_F\mu_B\bar{B_n}_z(\mathbf{r})\right|$
associated with the $z$-component of the n-th rf field (in the frame
aligned with the vector $\mathbf{B}_S(\mathbf{r})$) is small
compared to the rf frequency. In this case we can approximate
\begin{eqnarray}
J_{i}\left[\gamma_n(\mathbf{r})\right]\approx\left\{\begin{array}{c}
                                         1\quad\text{if}\quad i=0 \\
                                         0\quad\text{if}\quad i\neq 0
                                       \end{array}\right..
                                       \label{eq:bessel_small_argument}
\end{eqnarray}
This allows us to replace
$U_{T}^\dagger(\mathbf{r},t)\mathbf{F}U_{T}(\mathbf{r},t)$ in
equation (\ref{eq:final_hamiltonian}) by $\mathbf{F}$ itself,
yielding a major simplification. As long as $\omega \gg
\omega_{L\parallel}$ the oscillating motion of the field amplitude
will dominate the Lamor precession around the magnetic field
vector. Later on we will see that even if $\omega \gg
\omega_{L\parallel}$ is not strictly satisfied, i.e. both
frequencies differ only by one order of magnitude, the above
approximation works surprisingly well.

In the next step we utilize the so-called
rotating-wave-approximation which essentially consists of the
negligence of all terms that oscillate rapidly with a frequency
$2\omega$. Accounting for equation (\ref{eq:bessel_small_argument})
this give rise to the Hamiltonian
\begin{eqnarray}
H_\text{RF}&=&\frac{1}{2M}\left[\mathbf{p}+\mathbf{A}(\mathbf{r},t)\right]^2+g_F\mu_B\left[\left|\mathbf{B}_S(\mathbf{r})\right|
  -\frac{g_F}{|g_F|}\frac{\hbar\omega}{g_F\mu_B}\right]F_z\label{eq:RWA_hamiltonian}\\
  &&+\frac{g_F\mu_B}{2}\sum_n\left[\left\{\mathfrak{R}_z\left[-\zeta_n\right]\bar{\mathbf{B}}_n(\mathbf{r})\right\}\cdot
  \mathbf{F}-\bar{B_n}_z(\mathbf{r})F_z]\right]\nonumber\\
  &=&\frac{1}{2M}\left[\mathbf{p}+\mathbf{A}(\mathbf{r},t)\right]^2+g_F\mu_B\mathbf{B}_\text{eff}(\mathbf{r})\cdot
\mathbf{F}\nonumber
\end{eqnarray}
The dynamics of the spin particle is thus effectively determined
by a Hamiltonian which consists of two gauge potentials
$\mathbf{A}(\mathbf{r},t)$ and $\Phi(\mathbf{r},t)$ and a coupling
to an effective magnetic field whose components read
\begin{eqnarray}
 {B_\text{eff}}_x(\mathbf{r})&=&\frac{1}{2}\sum_n\left\{\mathfrak{R}_z\left[-\zeta_n\right]\bar{\mathbf{B}}_n(\mathbf{r})\right\}_x\\
  {B_\text{eff}}_y(\mathbf{r})&=&\frac{1}{2}\sum_n\left\{\mathfrak{R}_z\left[-\zeta_n\right]\bar{\mathbf{B}}_n(\mathbf{r})\right\}_y\\
  {B_\text{eff}}_z(\mathbf{r})&=&\left|\mathbf{B}_S(\mathbf{r})\right|
  -\frac{\hbar\omega}{|g_F|\mu_B}.
\end{eqnarray}
Apart from the gauge potentials the new time-independent
Hamiltonian has acquired the same form as our initial Hamiltonian
(\ref{eq:initial_hamiltonian}) but with a static effective
magnetic field. The main advantage is now that
$\mathbf{B}_\text{eff}$ is not to satisfy Maxwell's equations
since it is no true magnetic field. \textit{This is exactly the
reason why combined static and rf fields permit the design of a
much larger variety of traps than it is possible by solely using
static fields.} Moreover, the dependence of the $\zeta_n$ on the
sign of the atomic $g$-factor enables one to realize
state-dependent potentials as outlined in Ref.
\cite{Lesanovsky06}.

To obtain the adiabatic rf potentials we diagonalize the last term
of the Hamiltonian (\ref{eq:RWA_hamiltonian}) by applying the
transformation
\begin{eqnarray}
  U_F(\mathbf{r})=\exp\left[-i{F_z}\tilde{\alpha}(\mathbf{r})\right]\exp\left[-i{F_y}\tilde{\beta}(\mathbf{r})\right]\label{eq:unitary_trafo_2}.
\end{eqnarray}
This transformation is similar to the one given by equation
(\ref{eq:unitary_trafo_1}) but with the rotation angles
$\tilde{\alpha}(\mathbf{r})$ and $\tilde{\beta}(\mathbf{r})$ now
being defined as
\begin{eqnarray}
\tilde{\alpha}(\mathbf{r})=\arctan\left[\frac{{B_\text{eff}}_y(\mathbf{r})}{{B_\text{eff}}_x(\mathbf{r})}\right]&\quad\text{and}\quad&
\tilde{\beta}(\mathbf{r})=\arctan\left[\frac{\sqrt{{B_\text{eff}}_x^2(\mathbf{r})+{B_\text{eff}}_y^2(\mathbf{r})}}{{B_\text{eff}}_z(\mathbf{r})}\right].
\end{eqnarray}
Essentially this leads to the Hamiltonian
\begin{eqnarray}
  H_\text{final}=\frac{1}{2M}\left[\mathbf{p}+\mathbf{A^\prime}(\mathbf{r},t)\right]^2+g_F\mu_B|\mathbf{B}_\text{eff}(\mathbf{r})|F_z
\end{eqnarray}
with the new gauge field $\mathbf{A^\prime}(\mathbf{r},t)$ being
defined according to equation (\ref{eq:gauge_potentials}) but with
the transformation $U(\mathbf{r},t)$ being replaced by
$U^\prime(\mathbf{r},t)=U_S(\mathbf{r})U_{T}(\mathbf{r},t)U_{F}(\mathbf{r})$.

Finally we perform the adiabatic approximation which, just like
for static magnetic traps \cite{Folman02,Bill06}, essentially
consists of neglecting the gauge potential
$\mathbf{A^\prime}(\mathbf{r},t)$. This yields the adiabatic
Hamiltonian
\begin{eqnarray}
    H_\text{ad}=\frac{\mathbf{p}^2}{2M}+g_F\mu_B|\mathbf{B}_\text{eff}(\mathbf{r})|F_z=\frac{\mathbf{p}^2}{2M}+V_\text{ad}(\mathbf{r})
\end{eqnarray}
with dressed adiabatic radio frequency induced potentials
\begin{eqnarray}
  V_\text{ad}(\mathbf{r})=m_Fg_F\mu_B\left[\left(\left|\mathbf{B}_S(\mathbf{r})\right|
  -\frac{\hbar\omega}{|g_F\mu_B|}\right)^2+{B_\text{eff}}_x^2(\mathbf{r})+{B_\text{eff}}_y^2(\mathbf{r})\right]^\frac{1}{2}.\label{eq:adiabatic_potentials}
\end{eqnarray}
These potentials are time-independent and turn out to be extremely
versatile in terms of their applicability. Of course the validity of
the approximation utilized in this section has to be ensured.
However, it is hard to discuss this in general and has therefore to
be checked for the individual field setup. In the next section we
will restrict our considerations to a specific case.

\section{Two-dimensional rf potentials} \label{sec:ioffe_plus_rf}
Let us now demonstrate how a very simple and easy to build
experimental setup can give rise to extremely versatile dressed
adiabatic potentials that can be used for building atom optical
elements such as beam-splitters and interferometers.

To this end we consider two-dimensional potentials which are
generated from a static Ioffe trap whose field is given by
\begin{eqnarray}
\mathbf{B}_S(\mathbf{r})=Gx\mathbf{e}_x-Gy\mathbf{e}_y+B_I\mathbf{e}_z.
\end{eqnarray}
This two-dimensional trapping field shall be superimposed by two
homogeneous and mutually orthogonal radio-frequency fields
\begin{eqnarray}
  \mathbf{B}_1(\mathbf{r})=\frac{B_\text{RF}}{\sqrt{2}}\mathbf{e}_x
  &\textrm{and} &
  \mathbf{B}_2(\mathbf{r})=\frac{B_\text{RF}}{\sqrt{2}}\mathbf{e}_y.\label{eq:linear_rf_fields}
\end{eqnarray}
We choose the corresponding phase angles to be  $\delta_1=0$ and
$\delta_2=\delta$. After inserting this fields the Hamiltonian
(\ref{eq:initial_hamiltonian}) becomes
\begin{eqnarray}
  H_\text{2RF}(t)=\frac{\mathbf{p}^2}{2M}+g_F\mu_B\left[\left[Gx+\frac{B_\text{RF}}{\sqrt{2}}\cos\omega t\right]F_x-
  \left[Gy-\frac{B_\text{RF}}{\sqrt{2}}\cos\left(\omega
  t-\delta\right)\right]F_y+B_IF_z\right].
  \label{eq:hamiltonian_2rf_ioffe}
\end{eqnarray}
The goal of the following discussion is two-fold: Firstly we will
calculate the dressed adiabatic potentials arising from the
spin-field coupling term of the Hamiltonian
(\ref{eq:hamiltonian_2rf_ioffe}). We will point out the flexibility
of this rather simple field configuration which enables one to
create a number of atom optical elements. Secondly we want to
compare our results to a numerically exact wave packet propagation
that is governed by the Schr\"odinger equation
\begin{eqnarray}
   i\hbar\partial_t\left|\Psi(t)\right>=H_\text{2RF}(t)\left|\Psi(t)\right>.\label{eq:schroedinger_equation}
\end{eqnarray}

\subsection{Adiabatic potentials}
In order to calculate the dressed adiabatic potentials we need to
construct and apply the unitary transformation
(\ref{eq:unitary_trafo_1}) to the Hamiltonian
(\ref{eq:initial_hamiltonian}). With the rf fields
(\ref{eq:hamiltonian_2rf_ioffe}) the corresponding angles read
\begin{eqnarray}
\alpha(\mathbf{r})=\arctan\left[-\frac{y}{x}\right]&\quad\text{and}\quad&
\beta(\mathbf{r})=\arctan\left[\frac{G}{B_I}\sqrt{x^2+y^2}\right].
\end{eqnarray}
Employing cylindrical coordinates the rotation matrices given by
equation (\ref{eq:rotation_of_bs}) read
\begin{eqnarray}
  \mathfrak{R}_y\left[\beta(\mathbf{r})\right]
  \mathfrak{R}_z\left[\alpha(\mathbf{r})\right]=
  \left(
    \begin{array}{ccc}
      \frac{B_I}{\sqrt{B_I^2+G^2\rho^2}} & 0 & -\frac{G\rho}{\sqrt{B_I^2+G^2\rho^2}} \\
      0 & 1 & 0 \\
      \frac{G\rho}{\sqrt{B_I^2+G^2\rho^2}} & 0 & \frac{B_I}{\sqrt{B_I^2+G^2\rho^2}} \\
    \end{array}
  \right)
  \left(
  \begin{array}{ccc}
  \cos\phi & -\sin\phi & 0 \\
  \sin\phi & \cos\phi & 0 \\
  0 & 0 & 1 \\
  \end{array}
   \right)
\end{eqnarray}
with the polar angle $\phi$ and the radius $\rho$. According to
equation (\ref{eq:adiabatic_potentials}) we then find the following
dressed adiabatic potential:
\begin{eqnarray}
  \frac{V_\text{ad}(\mathbf{r})}{m_Fg_F\mu_B}=\left[\left(\left|\mathbf{B}_S(\mathbf{r})\right|-\frac{\hbar\omega}{|g_F\mu_B|}\right)^2
  +\frac{B^2_\text{RF}}{4}\left(1+\frac{B_I\sin\delta}{\left|\mathbf{B}_S(\mathbf{r})\right|}+
  \frac{G^2\rho^2}{2\left|\mathbf{B}_S(\mathbf{r})\right|^2}\left(\cos\delta\sin
  (2\phi)-1\right)\right)\right]^\frac{1}{2}.\label{eq:general_ioffe_plus_rf}
\end{eqnarray}
For $\cos\delta>0$ we find the minima and maxima of the potential at
$\phi_\text{min}=\frac{3}{4}\pi,\frac{7}{4}\pi$ and
$\phi_\text{max}=\frac{1}{4}\pi,\frac{5}{4}\pi$, respectively. If
$\cos\delta<0$ the positions of the minima and maxima simply
exchange. Assuming $\rho\ll B_\text{I}/G$ the radial position of
these extrema evaluates to
\begin{eqnarray}
\rho_0=\frac{1}{2G}\sqrt{B_\text{RF}^2(1-\cos\delta\sin(2\phi)+\sin\delta)-2B^2_\text{C}}\label{eq:splitting_distance}
\end{eqnarray}
with the critical field strength
$B_\text{C}=2\sqrt{B_\text{I}\frac{\hbar\triangle}{|g_F \mu_B|}}$
and the detuning $\hbar\triangle=|g_F \mu_B|B_\text{I}-\hbar\omega$.
Hence for $\cos\delta>0$ and
$B_\text{RF}<\sqrt{\frac{2}{1+\cos\delta+\sin\delta}}B_\text{C}$ or
$\cos\delta<0$ and
$B_\text{RF}<\sqrt{\frac{2}{1-\cos\delta+\sin\delta}}B_\text{C}$
solely a single minimum with respect to the radial coordinate can be
achieved. For $\delta=\frac{3}{2}\pi$ in any case only a single
minimum is found.

We now inspect the condition (\ref{eq:parallel_lamor_frequency}).
For the current setup we find
\begin{eqnarray}
  \gamma_1=\frac{g_F \mu_B B_\text{RF}\,G\rho}{\sqrt{2}\hbar\omega\,\left|\mathbf{B}_S(\mathbf{r})\right|}\cos\phi
  &\text{and} & \gamma_2=\frac{g_F \mu_B
  B_\text{RF}\,G\rho}{\sqrt{2}\hbar\omega\,\left|\mathbf{B}_S(\mathbf{r})\right|}\sin\phi
\end{eqnarray}
which can in case of a resonant rf field ($g_F \mu_B
  B_\text{RF}\approx \hbar\omega$) be approximated by
\begin{eqnarray}
   \gamma_{1,2}\approx\frac{G\rho}{2\left|\mathbf{B}_S(\mathbf{r})\right|}\approx
\frac{1}{2}\frac{G\rho}{B_I}-\frac{1}{4}\left[\frac{G\rho}{B_I}\right]^3.
\end{eqnarray}
Thus if the splitting distance $\rho_0$ is kept small and at the
same time the Ioffe field strength high the validity of the
condition $\gamma_{1,2}\ll 1$ can be ensured.

\subsection{Numerical wave packet propagation}
To obtain exact results and in particular to estimate the quality of
the adiabatic approach we will now perform a numerically exact wave
packet propagation. Consider the wave packet $\left|\Psi(t)\right>$
which we decompose according to
\begin{eqnarray}
\left|\Psi(t)\right>=\sum_{n m m_F}
c_{nmm_F}(t)\left|m,n\right>\left|m_F\right>
\label{eq:quantum_state}
\end{eqnarray}
where the functions $\left|m,n\right>$ are the orthonormal
eigenfunctions of a two-dimensional harmonic oscillator in Cartesian
coordinates
\begin{eqnarray}
\left|m,n\right>=\frac{\sqrt[4]{M^2\omega_x
\omega_y}}{\sqrt{2^{m+n}\pi\hbar\,
m!\,n!}}\,e^{-\frac{M}{2\hbar}\left(\omega_x x^2+\omega_y
y^2\right)
}H_m\left(\sqrt{\frac{M\omega_x}{\hbar}}x\right)H_n\left(\sqrt{\frac{M\omega_y}{\hbar}}y\right).
\label{part2b:eq:basis_set_cart}
\end{eqnarray}
The frequencies $\omega_x$ and $\omega_y$ can be regarded as
parameters which can be adapted in order to improve the convergence
of the numerical propagation \cite{Bill06}. To cover the spin space
dynamics we utilize the spinor-orbitals $\left|m_F\right>$,
respectively. Inserting the state (\ref{eq:quantum_state}) into the
Schr\"odinger equation (\ref{eq:schroedinger_equation}) and
multiplying by
$\left<m_F^\prime\right|\left<m^\prime,n^\prime\right|$ from the
left yields the set of ordinary differential equations
\begin{eqnarray}
i\hbar\partial_t c_{nmm_F}(t)=\sum_{n^\prime m^\prime m^\prime_F}
\left<m_F^\prime\right|\left<m^\prime,n^\prime\right|H_\text{2RF}(t)\left|n,m\right>\left|m_F\right>c_{n^\prime
m^\prime m^\prime_F}(t)
\end{eqnarray}
which can now be used to propagate the coefficients $c_{nmm_F}(t)$
in time. For all practical purposes this system of first order
differential equations has to be truncated yielding a set of
equations of finite dimension. Then it can be solved by using
standard numerical integration methods. In particular we have used a
Runge-Kutta integrator with adaptive stepsize provided by the
\textsc{Nag} library. This method is not norm-conserving. Thus we
have to ensure the conservation of the norm at any time step which
at the same time serves as a measure of the quality of the
propagation. During the propagation we solely ramp up the rf field
strength $B_\text{RF}$ form zero to its final value. All other
parameters remain unchanged. At $t=0$ and consequently
$B_\text{RF}=0$ the Hamiltonian (\ref{eq:hamiltonian_2rf_ioffe})
resembles that of a stationary Ioffe-Pritchard trap. For a
sufficiently large Ioffe field strength, i.e. $G\rho\ll B_I$, the
ground state of an atom of mass $M$ in this trap can be approximated
by \cite{Bill06}
\begin{eqnarray}
  \Psi_0(x,y)=\left<\mathbf{r}\mid\Psi_0\right>=\sqrt{\frac{M\omega}{\pi \hbar}}\exp\left[-\frac{M\omega}{2\hbar}\left(x^2+y^2\right)\right]\left|m_F\right>\label{eq:ground_state}
\end{eqnarray}
with the trap frequency
\begin{eqnarray}
  \omega=G\sqrt{\frac{g_F\mu_B m_F}{M B_I}}.
\end{eqnarray}
Thereby we assume the atom to be in a hyperfine sub-state with
$m_Fg_F>0$. For all numerical calculations presented in this work we
use the wave function (\ref{eq:ground_state}) as initial state, i.e.
$\left|\Psi(x,y,t=0)\right>=\left|\Psi_0(x,y)\right>$.
\begin{figure}[htb]\center
\includegraphics[angle=0,width=7cm]{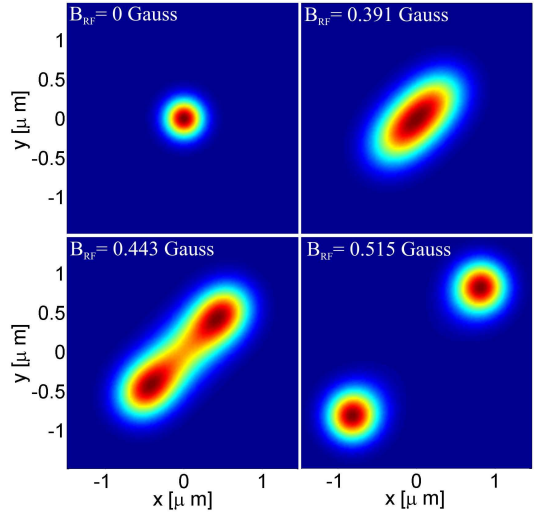}
\caption{Propagation of a wave packet from a single well into a
double well for $G=20\, \text{T/m}$, $B_I=0.75\, \text{Gauss}$,
$g_F=-\frac{1}{2}$, $m_F=-1$ and $\delta=\pi$. $B_\text{RF}$ is
linearly ramped from zero to $0.515\, \text{Gauss}$ within 7.6 ms.
The radio frequency is $\omega=2\pi\times 500 \text{kHz}$. Shown
is the probability density
$\left|\Psi(x,y)\right|^2$.}\label{fig:double_well}
\end{figure}
In figure \ref{fig:double_well} we present the propagation of an
atomic $^{87}\text{Rb}$ wave packet from the ground state in a
Ioffe-Pritchard trap into a double well. The relative phaseshift
between the two rf fields is $\pi$. The atom is supposed to be in
the upper branch of the $F=1$-manifold of the ground state, i.e.
$g_F=-\frac{1}{2}$ and $m_F=-1$. We ramp the amplitude linearly from
zero to $0.515\, \text{Gauss}$ over a period of 7.6 ms. All other
parameters remain at constant values $G=20\, \text{T/m}$,
$B_I=0.75\, \text{Gauss}$. The wave packet propagation reproduces
the results which one would expect from inspecting the adiabatic
potentials. For $\delta=\pi$ the potential minima are located on the
line defined by $x=y$. According to equation
(\ref{eq:splitting_distance}) the final splitting distance is
$\rho_0=1.4\, \mu m$ which is also quite well reproduced. For this
value of $\rho_0$ the parameters $\gamma_{1,2}$ evaluate to
approximately $0.25$. Hence, although the condition $\gamma_{1,2}\ll
1$ is not strictly satisfied the adiabatic description appears to
work quite well.

\begin{figure}[htb]\center
\includegraphics[angle=0,width=7cm]{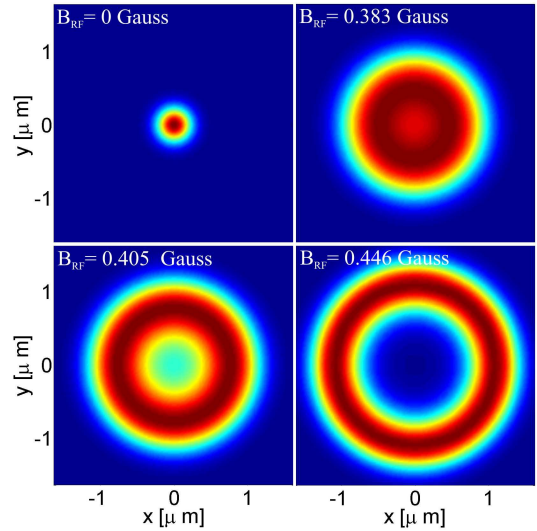}
\caption{Propagation of a wave packet from a single well into a ring
potential for $G=20\, \text{T/m}$, $B_I=0.75\, \text{Gauss}$,
$g_F=-\frac{1}{2}$ and $m_F=-1$. $B_\text{RF}$ is linearly ramped
from zero to $0.446\, \text{Gauss}$ within 7.6 ms. The radio
frequency is $\omega=2\pi\times 500 \text{kHz}$. Shown is the
probability density $\left|\Psi(x,y)\right|^2$.}\label{fig:ring}
\end{figure}
We now consider a relative phase shift of $\delta=\frac{\pi}{2}$
between the two rf fields. In this case the potential
(\ref{eq:general_ioffe_plus_rf}) becomes independent of the polar
angle $\phi$ and hence rotationally symmetric. For our numerical
wave packet propagation we start again in the ground state
(\ref{eq:ground_state}) and at zero rf amplitude. Subsequently we
ramp the rf amplitude to $0.446\, \text{Gauss}$ within 7.6 ms.
This is done linearly at a constant rf frequency of
$\omega=2\pi\times 500 \text{kHz}$. The resulting probability
density is depicted in figure \ref{fig:ring}. The initially
Gaussian distribution is isotropically deformed until a
ring-shaped wave function emerges. For the parameters given the
final ring radius evaluates according to equation
(\ref{eq:splitting_distance}) to $\rho_0= 1.08\, \mu m$ which is
in very good agreement with the numerical simulation (one finds
$\gamma_{1,2}=0.16$).

\section{Interference experiments in an rf induced double-well} \label{sec:interference}
The excellent performance of rf induced potentials for conducting
interference experiments has been both demonstrated experimentally
\cite{Schumm05} and studied theoretically \cite{Lesanovsky06}. It
was shown that by using the rf scheme splitting distances of only a
few microns can be achieved. Such small splittings cannot be
observed directly by means of absorption imaging. However, by
switching off all magnetic fields (which is assumed to happen
instantaneously) and after waiting a sufficiently long period of
time the structure of the expanded cloud can be resolved. In case of
an initially split cloud usually an interference pattern is observed
\cite{Schumm05}.

\begin{figure}[htb]\center
\includegraphics[angle=0,width=6cm]{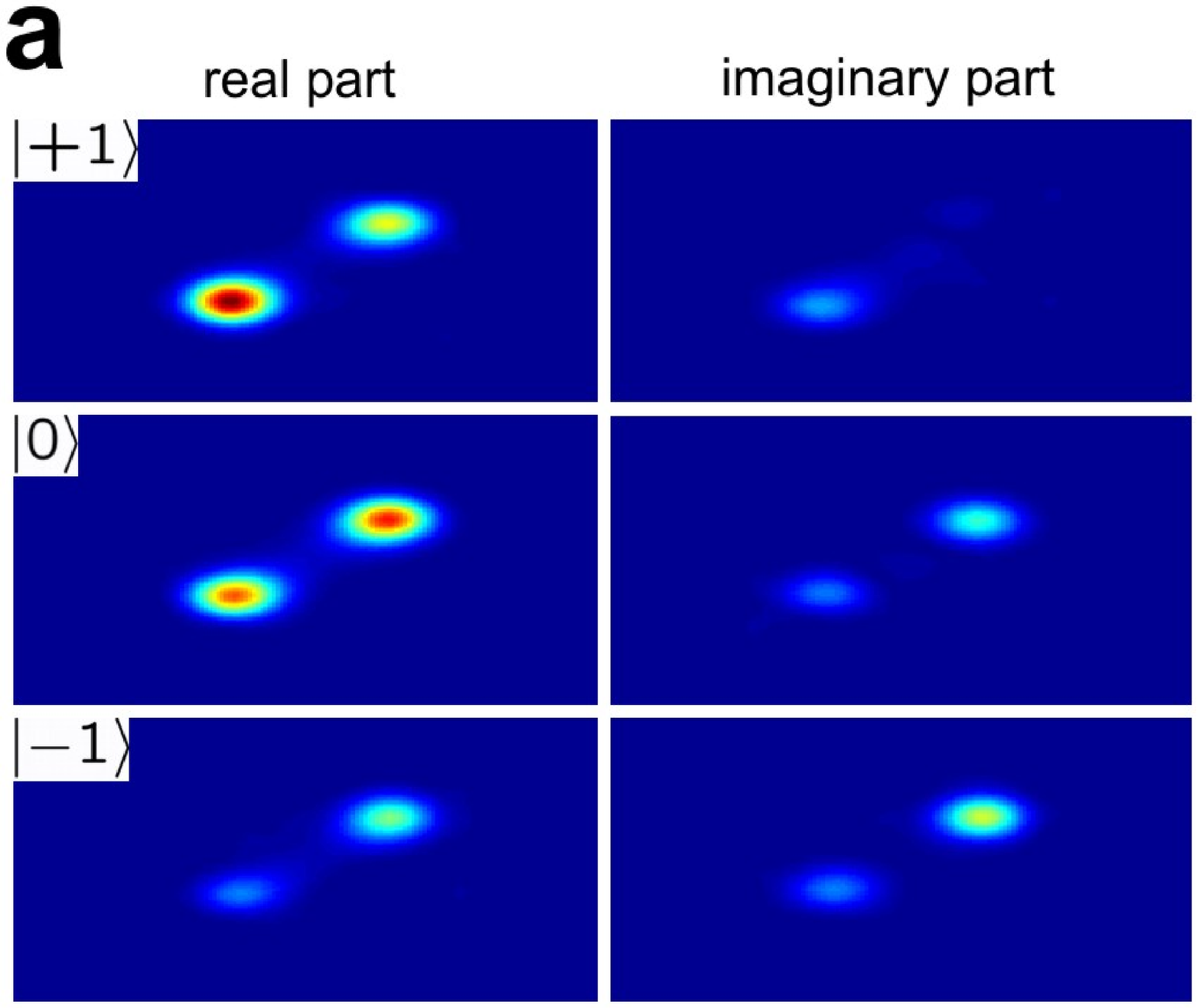}
\includegraphics[angle=0,width=7cm]{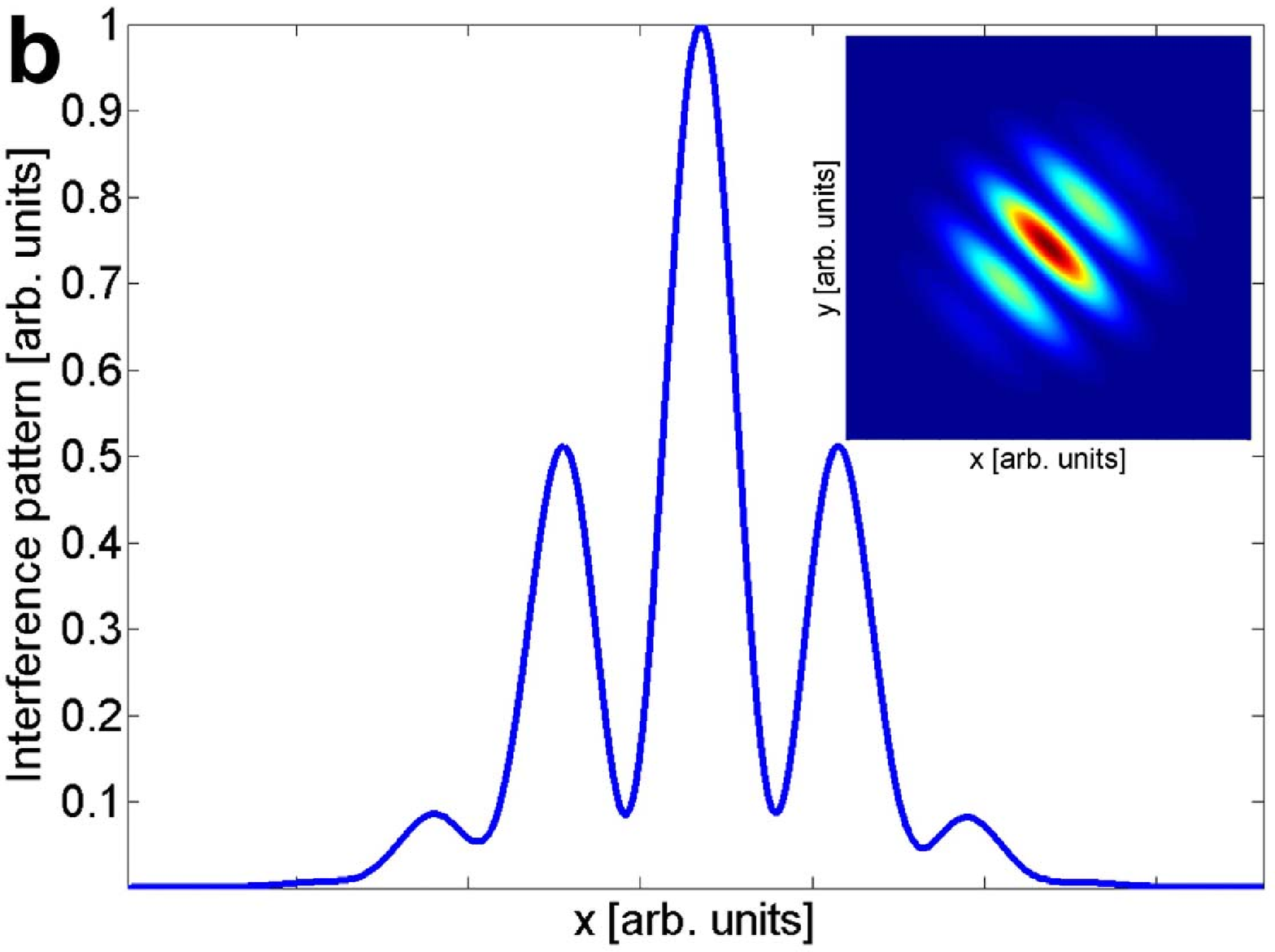}
\caption{\textbf{a}: Real and imaginary parts of the wave function
components which occupy the spinor orbitals $\left|m_s\right>$. The
state shown is the final state of the propagation presented in
figure \ref{fig:double_well}. For each plot the same colormap was
used. \textbf{b}: Cut along $x=y$ through the interference pattern
after a time-of-flight experiment. The interference contrast is
significantly smaller than $100\,\%$. This is essentially a
consequence of the different spin states of the two interfering
clouds.} \label{fig:dw_interference}
\end{figure}

We will now discuss how the nature of the rf potentials gives rise
to a reduction of the interference contrast. Assuming a free
propagation of the initial state being characterized by the wave
function $\Psi(\mathbf{x},t_0)$ we find the probability density
after a time $t+t_0$
\begin{eqnarray}
  \left|\Psi(\mathbf{x},t+t_0)\right|^2=\frac{1}{(2\pi)^{2d}}\left[\frac{m\pi}{2\hbar
  |t-t_0|}\right]^d\left|\mathfrak{F}\left\{\exp\left(\frac{im\mathbf{x}^2}{2 \hbar
  |t-t_0|}\right)\times\Psi(\mathbf{x},t_0)\right\}\left(\frac{m\mathbf{x}}{\hbar|t-t_0|}\right)\right|^2\label{eq:time_evolution}
\end{eqnarray}
with $d$ being the number of spatial dimensions considered and
$\mathfrak{F}\left\{g(\mathbf{x}^\prime)\right\}(\mathbf{x})$ being
the Fourier transform of the function $g(\mathbf{x}^\prime)$
evaluated at the position $x$. If $|t-t_0|$ becomes large and the
initial state is localized the position dependent phase factor
$\exp\left(\frac{im\mathbf{x}^2}{2 \hbar
  |t-t_0|}\right)$ becomes approximately uniform over the
extension of $\Psi(\mathbf{x},t_0)$ and can thus be taken out from
the argument and put in front of the Fourier integral. In this case
the the probability density after the time-of-flight period is
simply the Fourier transform of $\Psi(\mathbf{x},t_0)$. Since
$\Psi(\mathbf{x},t_0)$ is a spinor wave function the Fourier
transform is to be taken of each spinor component separately. Hence
one would expect the occupation of the individual spinor orbitals to
have an effect on the interference pattern. For the purpose of
demonstration we now consider the final state of the wave packet
propagation which is shown in figure \ref{fig:double_well}. In
figure \ref{fig:dw_interference}a we present the spin decomposition
of this particular state.
The probability amplitude seems to be randomly distributed over the
three spinor orbitals. None of the orbitals shows a spatially
symmetric occupation like the squared modulus of the wave function
which has interesting consequences for the interference contrast.
From the squared modulus of the final state's wave function (see
figure \ref{fig:double_well}) one would expect to achieve $100\,\%$
of contrast in the interference pattern, i.e. $\min
\left|\Psi(\mathbf{x},t+t_0)\right|^2=0$. The actual calculation of
the interference pattern (\ref{fig:dw_interference})b, however,
yields a contrast of about $90\,\%$. This modification of the
interference pattern by the spatially asymmetric occupation of the
spinor orbitals will be subject of the following discussion.

\begin{figure}[htb]\center
\includegraphics[angle=0,width=5cm]{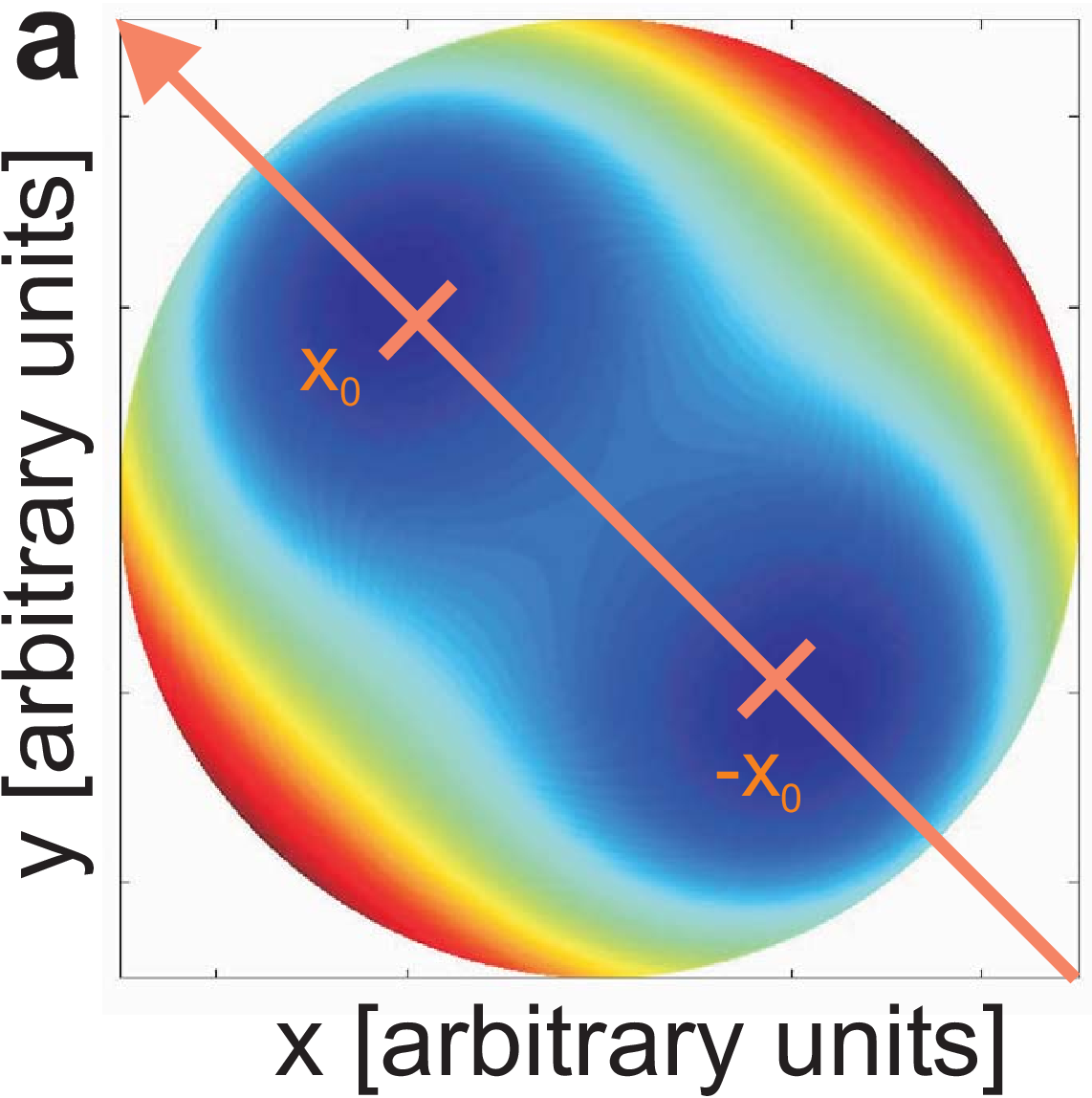}\qquad\qquad
\includegraphics[angle=0,width=6cm]{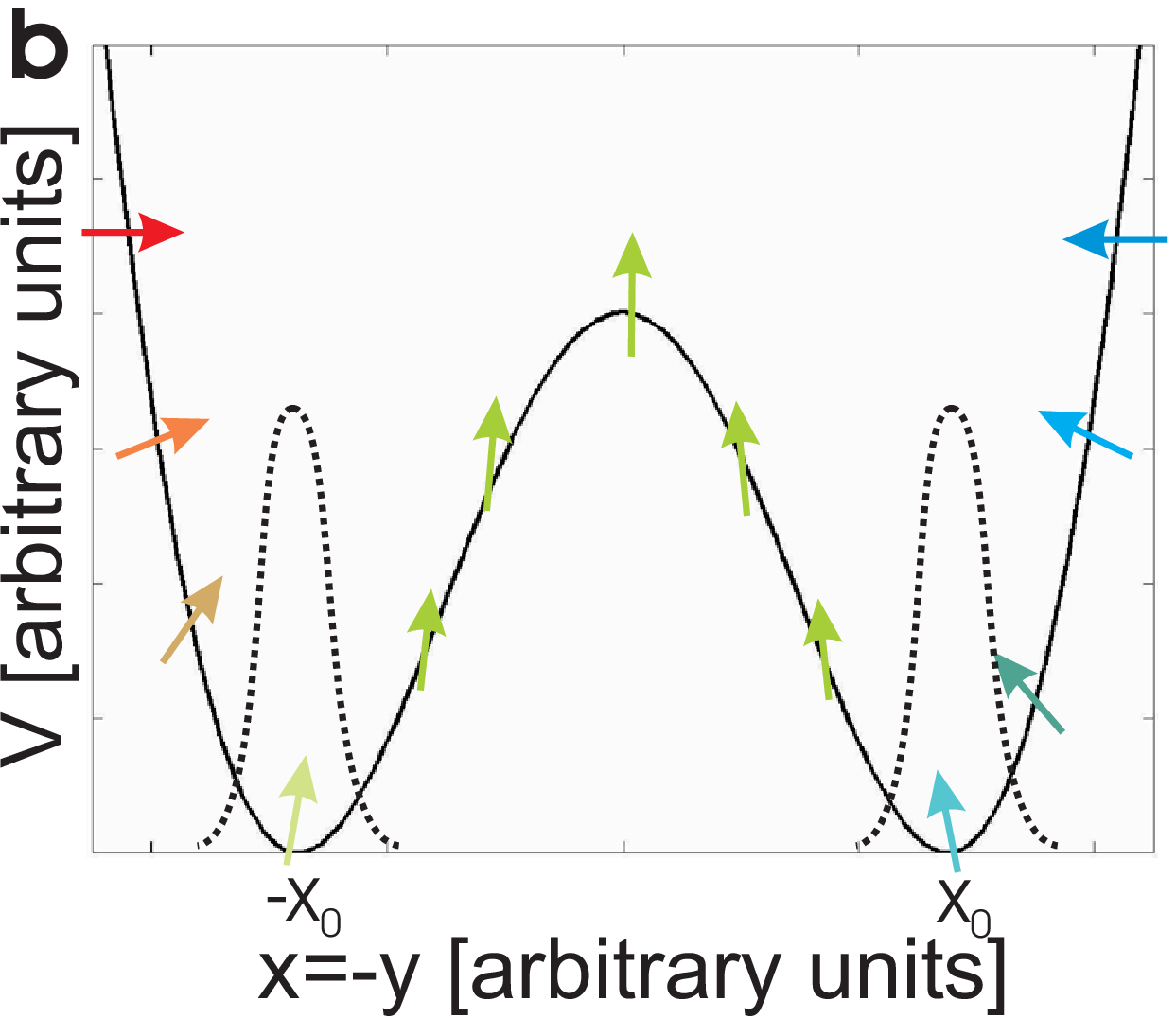}
\caption{\textbf{a}: Two-dimensional plot of the double well
potential (\ref{eq:general_ioffe_plus_rf}) for $\delta=0$. In order
to study the dependence of the interference contrast on the
splitting distance $2x_0$ we consider only the 1-dimensional motion
along the indicated axis. \textbf{b}: Sketch of the spin orientation
in an rf double-well potential along the axis being shown in
\textbf{a}. The spin orientation is assumed to follow adiabatically
the local direction of the effective magnetic field. Hence the
interference contrast of two interfering clouds depends in general
on the relative orientation of the two spin vectors.}
\label{fig:1d_approximation}
\end{figure}

We focus on the rf double-well which is described by equation
(\ref{eq:general_ioffe_plus_rf}) if $\delta=0$ or $\pi$. We consider
the interference pattern that emerges if two completely separated
clouds which are located in the two wells interfere. For the purpose
of illustration we consider the one-dimensional dynamics along the
axis which is defined through the position of the double-well minima
$\pm\mathbf{x}_0$ (see figure \ref{fig:1d_approximation}a). We
further model our initial state according to
\begin{eqnarray}
  \Psi(\mathbf{x},t=0)=\frac{1}{\sqrt{2}}\left[\delta(x-x_0)\left|x_0\right>_\text{S}+\delta(x+x_0)\left|-x_0\right>_\text{S}\right]\label{eq:model_state}.
\end{eqnarray}
The spatial part of the two separated wave functions is described by
delta functions. The spin is accounted for by the spinor orbitals
$\left|x\right>_\text{S}$. For the Fourier transform of the state
(\ref{eq:model_state}) one obtains
\begin{eqnarray}
  \Psi(k)=\frac{1}{\sqrt{2}}\left[\exp\left(ikx_0\right)\left|x_0\right>_\text{S}+\exp\left(-ikx_0\right)\left|-x_0\right>_\text{S}\right].
\end{eqnarray}
Hence, according to equation (\ref{eq:time_evolution}) the
probability density after a sufficiently long waiting period
evaluates to
\begin{eqnarray}
 |\Psi(\mathbf{x},t)|^2&\approx&\text{const}\times\left[2+\exp\left(-2i\frac{mx}{\hbar t}x_0\right)\left<x_0\mid-x_0\right>_\text{S}+\exp\left(2i\frac{mx}{\hbar t}x_0\right)\left<-x_0\mid
  x_0\right>_\text{S}\right]\nonumber\\
  &\propto& 1+\mathrm{Re}\left\{\exp\left(i\,k_0 x\right)\left<-x_0\mid
  x_0\right>_\text{S}\right\}\label{eq:contrast}
\end{eqnarray}
with $k_0=2\frac{mx_0}{\hbar t}$. If the shape of the spinor wave
function was independent of the spatial position $x_0$ one would
observe an interference pattern with $100\, \%$ contrast and a
period of $\frac{2\pi}{k_0}$. In general, however, we find
$\left|\left<-x_0\mid x_0\right>_\text{S}\right|\leq 1$ which is
simply the result of the adiabatic alignment of the spin along the
effective magnetic field vector which is illustrated in figure
\ref{fig:1d_approximation}b. This means that the spin states of the
two initial, now interfering, fragments were not identical. It is
easily seen this effect to lead to a reduction of the interference
contrast and to a variation of the global phase of the interference
pattern.

To calculate the quantity $\left<-x_0\mid x_0\right>_\text{S}$ we
use the unitary transformations which we have employed to achieve
the adiabatic rf potentials. In the dressed frame the spinor wave
function is independent of the spatial coordinate and will be
characterized by $\left|\text{ad}\right>$. To obtain the spin wave
function in the lab frame $\left|x\right>_\text{S}$ we have to apply
the unitary transformations (\ref{eq:unitary_trafo_1}),
(\ref{eq:rotating_frame}), (\ref{eq:trafo_removes_omega_term}) and
(\ref{eq:unitary_trafo_2}). After that we receive
\begin{eqnarray}
  \left|x\right>_\text{S}=U_F(x)U_T U_R
  U_S(x)\left|\text{ad}\right>.
\end{eqnarray}
Hence we find
\begin{eqnarray}
  \left<-x_0\mid x_0\right>_\text{S}= \left<\text{ad}\right|U_S^\dagger(-x_0) U_R^\dagger U_T^\dagger U_F^\dagger(-x_0)U_F(x_0)U_T U_R
  U_S(x_0)\left|\text{ad}\right>
\end{eqnarray}
This expression simplifies considerably if we employ cylindrical
coordinates. With $x_0=(\rho_0,\phi=\frac{3\pi}{4})$ and
$-x_0=(\rho_0,\phi=\frac{7\pi}{4})$ we end up with
\begin{eqnarray}
  U_F^\dagger(-x_0)U_F(x_0)=\exp\left[i \arctan\left(\frac{G\rho_0}{B_I}\right)F_y\right]\exp\left[iF_z \pi\right]\exp\left[-i
  \arctan\left(\frac{G\rho_0}{B_I}\right)F_y\right].
\end{eqnarray}
We now neglect the transformation $U_T$. Its effect is expected to
be small since we consider the case $\gamma_n\ll 1$. Furthermore we
note that
\begin{eqnarray}
  U_S(x_0)=U_S^\dagger(-x_0)=\exp\left[-i\arctan\left(\zeta(\rho_0)\right)\, F_y\right]
\end{eqnarray}
with
\begin{eqnarray}
 \zeta(\rho_0)=\frac{B_\text{RF}B_I}
  {2|\mathbf{B}_S(\rho_0)|\left[|\mathbf{B}_S(\rho_0)|-\frac{\hbar\omega}{|g_F\mu_B|}\right]}
  \approx|g_F\mu_B|B_{RF}\left[\frac{1}{2\hbar\triangle}-\frac{\hbar\triangle+|g_F\mu_B|B_I}{4}\left[\frac{G\rho_0}{\hbar\triangle\,B_I}\right]^2\right].
\end{eqnarray}
In case of a resonant rf one hence finds as a zeroth order
approximation $\zeta(\rho_0)\approx \frac{B_{RF}}{2B_I}$. Putting
everything together we finally arrive at
\begin{eqnarray}
  \left<-x_0\mid x_0\right>_\text{S}=
  \left<\text{ad}\right|U_S^\dagger(x_0)\exp\left[i\pi\frac{B_I F_z-G \rho_0\cos(\omega t)F_x+G \rho_0\sin(\omega
  t)F_y}{|\mathbf{B}_S(\rho_0)|}\right]
  U_S^\dagger(x_0)\left|\text{ad}\right>.\label{eq:interference_contrast}
\end{eqnarray}
For a  $F=1/2$ particle in its trapped adiabatic state
$\left|\text{ad}\right>=\left|m_F=\frac{1}{2}\right>$ one finds
\begin{eqnarray}
  \left<-x_0\mid
  x_0\right>_\text{S}=\frac{1}{|\mathbf{B}_S(\rho_0)|}\left[i B_I-\frac{\zeta(\rho_0)}{\sqrt{1+\zeta^2(\rho_0)}}\,G \rho_0 \sin(\omega
  t)\right]. \label{eq:interference_contrast_spin12}
\end{eqnarray}
Apparently the overlap $\left<-x_0\mid x_0\right>_\text{S}$ is not
only a function of the relative displacement of the two atom clouds
but also a function of the time $t$. This, however, implies the
interference contrast to depend on the actual phase of the rf field.

Evaluating the interference term in equation (\ref{eq:contrast}) we
finally obtain
\begin{eqnarray}
\mathrm{Re}\left\{\exp\left(i\,k_0 x\right)\left<-x_0\mid
  x_0\right>_\text{S}\right\}=-\frac{B_I}{|\mathbf{B}_S(\rho_0)|}\sin(k_0 x)-\frac{\zeta(\rho_0)}{\sqrt{1+\zeta^2(\rho_0)}}\frac{G
  \rho_0}{|\mathbf{B}_S(\rho_0)|}
  \sin(\omega
  t)\cos(k_0 x). \label{eq:spin12_contrast}
\end{eqnarray}
\begin{figure}[htb]\center
\includegraphics[angle=0,width=7cm]{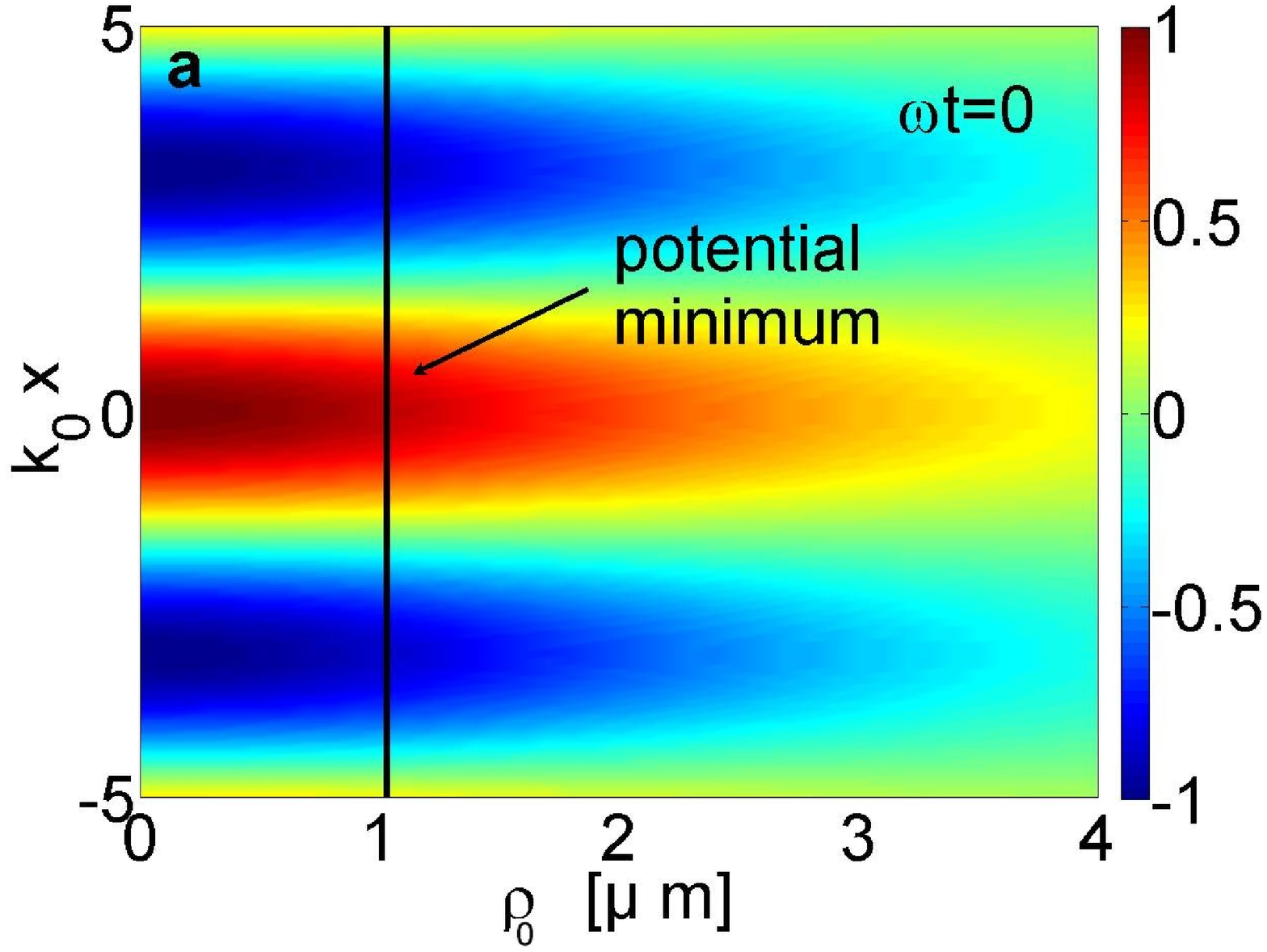}
\includegraphics[angle=0,width=7cm]{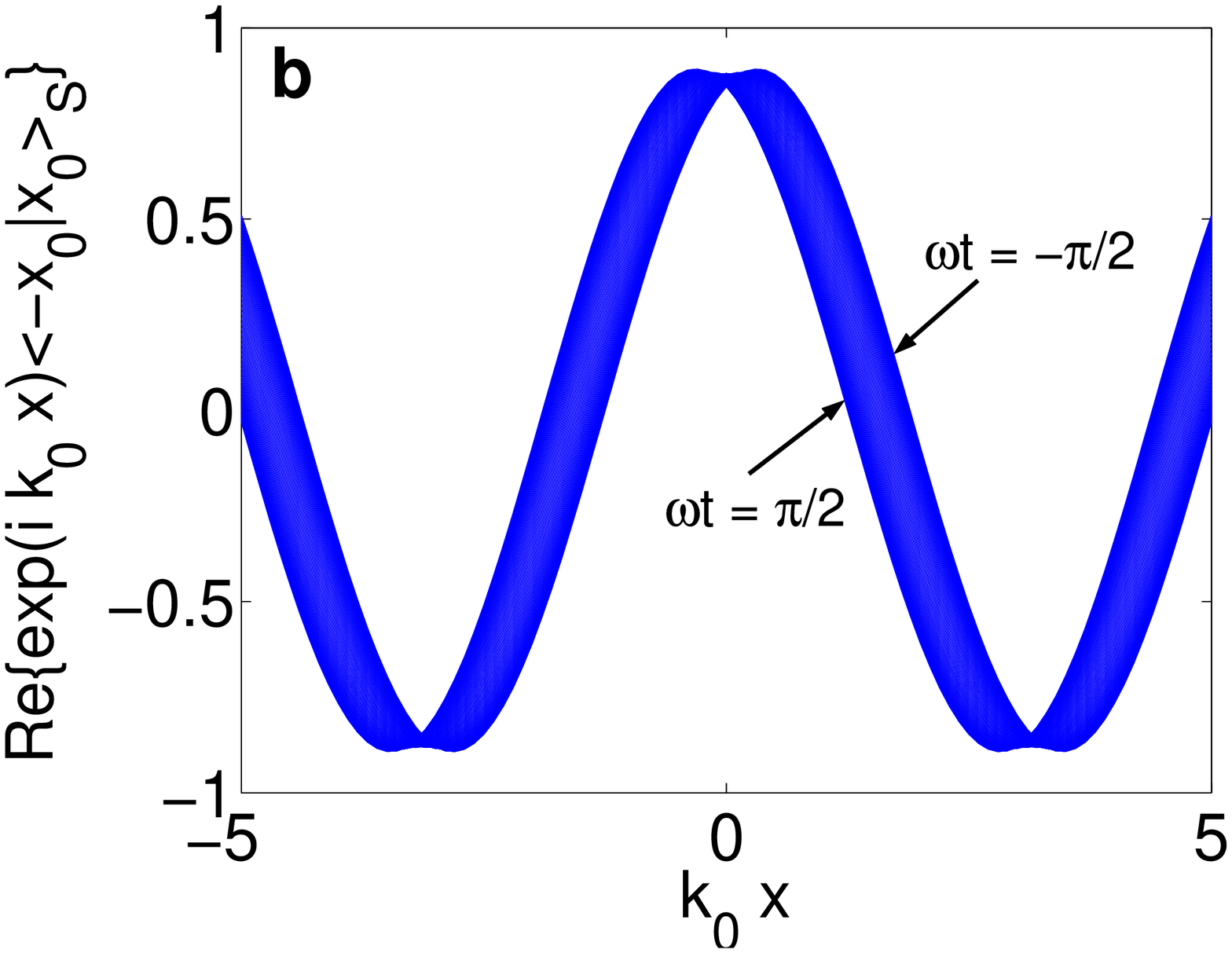}
\caption{\textbf{a}: Contrast pattern along the axis connecting the
two potential minima of the double-well (see figure
\ref{fig:1d_approximation}). The same parameters as in figure
\ref{fig:double_well} have been used. The snapshot is taken at
$\omega t=0$. The position of the potential minimum is indicated by
the vertical line. The interference contrast decreases as $\rho_0$
increases. For $\rho\rightarrow\infty$ the contrast goes to zero as
both wave packets occupy orthogonal spin states. \textbf{b}: Time
dependence of the interference if both wave packets are released
from the double-well minima. One can clearly recognize the
dependence of the interference pattern on the rf
phase.}\label{fig:contrast}
\end{figure}
For the calculation in case of $F>1/2$ the results given in
appendix \ref{txt:appendix_a} can be used. The general structure
of the interference term is similar for all cases. One encounters
a time-independent term which accounts for the static components
of the magnetic field and an oscillating one which gives rise to a
temporal modulation of the interference pattern. This can be
interpreted as the effect of the micro-motion of the atoms within
the adiabatic potentials similar to trapped ions in a Paul-trap.
Using equation (\ref{eq:spin12_contrast}) one finds a phase
variation of
\begin{eqnarray}
  \triangle \phi=2\arctan\left[\frac{\zeta(\rho_0)}{\sqrt{1+\zeta^2(\rho_0)}}\frac{G
  \rho_0}{|\mathbf{B}_S(\rho_0)|}\right]
  \approx 2\left[\frac{G\rho_0}{B_I}\right]\left[1+\left[\frac{2\hbar\triangle}{g_F\mu_B
  B_{RF}}\right]^2\right]^{-\frac{1}{2}}\label{eq:phase_wobbeling}
\end{eqnarray}
over one rf oscillation period. For a spin $1$ particle this rf
phase dependence of the interference pattern is shown in figure
\ref{fig:contrast}b. The right-hand side of equation
(\ref{eq:phase_wobbeling}) is the leading order term in
$\frac{G\rho_0}{B_I}$. Hence the phase oscillations can be
suppressed by keeping the ratio $\frac{G\rho_0}{B_I}$ small which is
in agreement with the previous assumption $G\rho\ll B_I$.

The above consideration assumes an immediate switch-off of all
external magnetic fields before the expansion of the matter wave.
However, for all practical purposes there is always a finite
switch off time. One might think of switching off the fields such
that finally all atoms are rotated into the same spin state but
the spatial shape of the matter wave remains unchanged. To succeed
in establishing such a sophisticated switching-off procedure seems
to be unlikely since at each spatial and temporal position the
atomic spin state had to be rotated differently.

\section{Summary and conclusion}\label{sec:conclusion}
We have presented the theoretical foundations for the description of
rf induced adiabatic potentials. Starting from a Hamiltonian that
takes into account the coupling of a single hyperfine manifold to an
external field we have carried out a number of unitary
transformations. After performing the rotating wave approximation
and neglecting the non-adiabatic couplings which emerge from the
transformed kinetic energy we have received the corresponding
adiabatic potential surfaces.

To demonstrate the power of this rf induced dressed adiabatic
potentials we have discussed a simple field configuration consisting
of a static Ioffe-Pritchard trap and two orthogonal homogeneous rf
fields. Our analytical calculations have shown that by tuning the
strength of the rf fields a smooth transition from a single well
into a double well is achievable. This transition can be easily
exploited to split a cloud of ultracold atoms. By introducing a
relative phase shift between the two rf fields furthermore a
transition from a double well to a ring potential can be performed.
For our considerations we have mainly focussed on
$^\text{87}\text{Rb}$ in the $F=1$ hyperfine ground state. For this
species the analytical results have been verified by a numerical
wave packet propagation which was conducted using a linear single
particle Schr\"odinger equation and the original Hamiltonian. For
typical experimental parameters these numerical results have been
shown to be in very good agreement with the ones obtained from the
analytic adiabatic approach.

Finally we have discussed interference experiments carried out in a
rf double-well potential. Since the trapping potential is a
consequence of a spin-field coupling the spin state of a trapped
atom cloud depend on its actual spatial position. This essentially
leads to a spatially asymmetric distribution of the atomic wave
function within the individual spinor orbitals and an inevitable
reduction of the interference contrast. Moreover, the rf field
imposes a high-frequency oscillation on the interference fringes
which is reminiscent of a micromotion. The magnitude of these
effects can be well controlled by appropriately tuning the
experimental parameters such as the Ioffe field strength and the
detuning.

\appendix
\section{$\left<-x_0\mid x_0\right>_\text{S}$ in case of $F\geq\frac{1}{2}$}\label{txt:appendix_a}
Overlap of the spin wave functions for atoms in the
$F={\frac{1}{2},1,\frac{3}{2},2}$-state calculated according to
equation (\ref{eq:interference_contrast}):
\begin{eqnarray}
  \left<-x_0\mid
  x_0\right>_\text{S}=\left[\frac{1}{|\mathbf{B}_S(\rho_0)|}
  \left( i B_I - \frac{\zeta(\rho_0)}{\sqrt{1+\zeta^2(\rho_0)}} G
\rho_0\sin(\omega t)\right)\right]^{2F}.\label{eq:spin1_contrast}
\end{eqnarray}
The atom is assumed to occupy the maximal stretched stated, i.e.
$m_F=F$. Like in the $F=\frac{1}{2}$-case (equation
(\ref{eq:interference_contrast_spin12})) we find a static and a
time-dependent part. Here, the latter has contribution of terms
that oscillate at the frequencies $\omega,..,2F\times\omega$.

\end{document}